\def\year{2019}\relax
\documentclass[letterpaper]{article} % DO NOT CHANGE THIS
\usepackage{aaai19}  % DO NOT CHANGE THIS
\usepackage{times}  % DO NOT CHANGE THIS
\usepackage{helvet} % DO NOT CHANGE THIS
\usepackage{courier}  % DO NOT CHANGE THIS
\usepackage[hyphens]{url}  % DO NOT CHANGE THIS
\usepackage{graphicx} % DO NOT CHANGE THIS
\usepackage{graphicx}  % DO NOT CHANGE THIS
\usepackage{booktabs}
\usepackage{makecell}
\usepackage{multirow}
\usepackage{dcolumn}
\usepackage[dvipsnames]{xcolor}
\usepackage{verbatim}
\usepackage{resizegather}
\newcommand{\newcite}[1]{\citeauthor{#1} \shortcite{#1}}
\newcommand{\myparagraph}[1]{\paragraph{#1}}

\nocopyright 

\begin{document}
\title{Smart, Responsible, and Upper Caste Only: \\Measuring Caste Attitudes through Large-Scale Analysis of Matrimonial Profiles}
\author{Ashwin Rajadesingan,\textsuperscript{\rm 1} Ramaswami Mahalingam,\textsuperscript{\rm 2} David Jurgens\textsuperscript{\rm 1}\\
\textsuperscript{\rm 1}School of Information, \textsuperscript{\rm 2}Department of Psychology\\
\texttt{\{arajades,ramawasi,jurgens\}@umich.edu}\\
University of Michigan, Ann Arbor
}
\maketitle

\begin{abstract}

Discriminatory caste attitudes currently stigmatize millions of Indians, subjecting individuals to prejudice in all aspects of life. Governmental incentives and societal movements have attempted to counter these attitudes, yet accurate measurements of public opinions on caste are not yet available for understanding whether progress is being made.  Here, we introduce a novel approach to measure public attitudes of caste through an indicator variable: openness to intercaste marriage.  Using a massive dataset of over 313K profiles from a major Indian matrimonial site, we precisely quantify public attitudes, along with differences between generations and between Indian residents and diaspora.
We show that younger generations are more open to intercaste marriage, yet attitudes are based on a complex function of social status beyond their own caste.  In examining the desired qualities in a spouse, we find that individuals open to intercaste marriage are more individualistic in the qualities they desire, rather than favoring family-related qualities, which mirrors larger societal trends away from collectivism.
Finally, we show that attitudes in diaspora are significantly less open, suggesting a bi-cultural model of integration.  
Our research provides the first empirical evidence identifying how various intersections of identity shape attitudes toward intercaste marriage in India and among the Indian diaspora in the US.

\end{abstract}

\section{Introduction}

Traditionally, marriages in India are considered as a union of two families. Typically, parents initiate and mediate the search for a spouse within kinship networks and ensure compatibility based on factors such as caste, education, affluence, horoscope and physical characteristics. These ``arranged'' marriages are usually endogamous with the bride and groom belonging to the same caste, reinforcing caste lines \cite{ambedkar2004castes}. However, recent studies show a decline in parental control in the matchmaking process, with more individuals reporting that they chose their spouse independently or at least jointly with their parents \cite{Allendorf_2013,Allendorf_Pandian_2016a}. Coupled with a greater openness to marry outside their caste, this increase in agency has the potential to alter the social fabric of the society, especially pertaining to caste relations. In this work, using data from over 313K profiles from a major Indian matrimonial site, we study social inclusion by examining partner preferences and factors that affect openness to intercaste marriage in India and the Indian diaspora in the US.  %

Previous research on intercaste marriage primarily relies on rich ethnographies \cite{fuller2008companionate,kannan1963intercaste}, qualitative interviews \cite{Dhar_2013}, surveys \cite{Goli_Singh_Sekher_2013,Ray_Chaudhuri_Sahai_2017} and audit studies \cite{Banerjee_Duflo_Ghatak_Lafortune_2013a,Ahuja_Ostermann_2016}. 
Unlike surveys which depend on self reported openness which may have a social desirability bias \cite{Krumpal_2013} or measuring incidence of intercaste marriages which may be different from actual openness, in our work, we use data from online matrimonial profiles where users specify their openness to intercaste marriage when they register.
As profiles contain rich demographic information, we can identify how various intersections of identity shape attitudes toward inter-caste marriage.

In Indian matrimonial sites, either individuals or their family can create a matrimonial profile, allowing us to study  generational differences in spousal preferences, and  by proxy, caste attitudes. 
Further, given these websites' popularity among Indian diaspora, through openness to intercaste marriage, we analyze how intercaste relations differ among diaspora in the US who live as a minority in a fundamentally different cultural context. 
These online profiles provide provide a unique real-time estimate of changing societal attitudes on caste and, because of their intensely personal nature, accurately reflect social preferences. 

Our work provides the following three main contributions.
First, we find younger generations are substantially more open to intercaste marriage when controlling for demographics and background (\S\ref{sec:big-regression}).
However, the demographic factors associated with openness are highly similar across generations, with the exception that higher education in children shows increased openness when compared with their family.
Moreover, we find clear geographic trends in acceptance to intercaste marriage, with South Indian states being less open. % 
Second, we show that individuals who are open to intercaste marriage are more likely to desire a spouse based on individual traits, whereas those not open to intercaste marriage emphasize family; these trends mirror the shift in Indian culture from collectivist to more individualist (\S\ref{section:text}).
Third, Indian diaspora show a lower acceptance to intercaste marriage compared to Indians in India (\S\ref{sec:diaspora}).
This result aligns with current theory \cite{mahalingam2013cultural} which contends that some diasporic individuals, because of their minority status, essentialize the notion of caste as a way maintaining cultural identity which results in resistance to intercaste marriage.

\section{Caste System}
\label{sec:caste}

Caste is a form of social hierarchy which assigns status to individuals based on birth. Originally based on the Hindu \textit{varna} system which grouped people based on their occupation, caste has evolved into a form of social identity based on birth which decides hierarchy and social order in society \cite{srinivas2003obituary}. The caste system divides people into groups and is organized based on perceived notions of ``purity'' and ``pollution'' \cite{dumont1980homo}. Brahmins hold the highest rank in this caste hierarchy, Other Forward Castes (OFC) are  non-Brahmin upper castes, Other Backward Castes (OBC) are castes considered to be lower economically and socially, and Scheduled Castes (SC or Dalits) and Scheduled Tribes (ST) are historically disadvantaged groups. The prejudice propagated by caste through centuries denied individuals in lower caste groups dignity and self respect \cite{kumar2010dalit} as well as access to education \cite{chauhan2008education}, job opportunities \cite{Banerjee_Knight_1985}, religious institutions \cite{tejani2013untouchable} and public resources such as water \cite{o2014public}. Despite increased awareness \cite{banerjee2015awareness}, government policies \cite{agrawal1991educational} and work of social reformers \cite{ambedkar2014annihilation}, caste based discrimination is still prevalent today. This form of discrimination greatly reduces social and economic mobility among these groups and increases inequality in society \cite{deshpande2011grammar,thorat2012blocked}. Caste discrimination is not limited to India, the Dalit diaspora in the US also face similar verbal, physical and workplace discrimination, with one in two Dalit respondents reporting that they fear being ``outed'' \cite{Zwick}. 

One of the primary institutions that reinforces caste lines is arranged marriages \cite{hopkins2017structure}, which are usually endogamous. Only 5\% of marriages in India are intercaste marriages where the bride and groom belong to different caste groups \cite{desai2015india}. Couples engaging in intercaste marriage face social seclusion \cite{parish1996hierarchy}, loss of family support \cite{Dhar_2013} and  violence \cite{chowdhry1997enforcing} even amounting to murder \cite{kumar2012public}. Nevertheless, there is increasing support for intercaste marriages through NGO networks \cite{kumar2012public}, activists \cite{rinker2013should} and government schemes that provide monetary support to encourage intercaste marriage. %  
Therefore, precise measurement of changing attitudes on caste is readily needed to highlight where increased pressure and incentives can be used to ultimately bring about an equitable society. %

\section{Matrimonial Profile Data}

The search for a spouse has given rise to multiple online matrimonial sites that host millions of profiles.
Crucially, unlike \textit{dating} websites, matrimonial sites are specifically tailored for marriage and include language prohibiting their use for casual relationships.
Like most matchmaking websites, individuals provide detailed information including photos, a free-form self description section, family details, age, location, caste, education level, income and current job.
However, unlike Western matchmaking platforms, these sites allow others (parents, siblings, other relatives and friends) to post on behalf of a person, thereby facilitating the traditional arranged marriage process in an online setting \cite{seth2011online}. 
Critically, the platform directly asks individuals to indicate their openness to intercaste marriage, which we  use as a proxy for their attitude toward caste.

Strong evidence indicates that the preferences expressed in these online matrimonial profiles accurately reflect offline personal attitudes in the Indian social context.  
Matrimonial sites in Indian communities facilitate---rather than replace---the traditional arranged marriage process, aiding the family or individual in selecting a spouse \cite{seth2011online,mathur2007s}.  These matrimonial sites help make connections that would have been traditionally made by extended family members but are potentially more difficult to make in modernized India due to geographic mobility \cite{agrawal2015cyber}.    
Indeed, user interviews indicate that these sites provide individuals with greater agency over the desired traits in a spouse, compared to the traditional parent-driven selection process, and therefore the expressed preferences are more closely mirror individual preferences \cite{titzmann2015matchmaking}.

\begin{table}[t]
    \centering
    \resizebox{0.47\textwidth}{!}{
    \begin{tabular}{r cc l}
         &\textbf{Self posted} & \textbf{Family posted} & \textbf{\textit{Total}}\\
        \cline{2-3} 
         \textbf{Male profiles} & 175,611 & 50,245 &  	225,856\\
         \textbf{Female profiles} & ~~39,039 & 48,413& 87,452\\
         \cline{2-3} 
         \textbf{\textit{Total}} & 214,650 & 98,658 & 313,308\\
    \end{tabular}
    }
    \caption{The stratified matrimonial profile dataset.}
    \label{tab:data}
\end{table}

\begin{table}[t]
    \centering
    \resizebox{0.48\textwidth}{!}{
    \begin{tabular}{l l c c }
         \textbf{Label} & \textbf{Caste Category} & \textbf{Pop. \%} & \textbf{Data \%} \\
         \hline
         Brahmin & Brahmin               & ~~5.2 & 16.62 \\
         OFC & Other Forward Castes       & 22.9 & 27.43\\
         OBC & Other Backward Castes & 40.5 & 39.38\\
         SC & Scheduled Castes      & 21.2 & 10.84\\
         ST & Scheduled Tribes      & ~~8.6  & ~~1.74\\
         Intercaste & Intercaste            & -  & ~~0.17\\
         Other & Other & ~~1.3 & ~~3.82\\
         
    \end{tabular}
    }
    \caption{Groupings of caste by IHDS \cite{desai2015india} for individuals self-identifying as Hindu and their prevalence in society and on our dataset. Pop. \% is the estimated percentage of population in India. Data \% is the percentage of users in our dataset stratified by caste. ``Others'' refers to individuals who do not self-identify with any caste or for whom there is no established mapping to caste category.  Data on intercaste populations are unavailable}
    \label{tab:castes}
\end{table}

\paragraph{Data Collection}
Data was collected from Shaadi.com, a leading Indian matrimonial site, by creating a profile to access public profile information available on the website. This experimenter-created profile included minimal information using, whenever possible, default profile attributes and site-provided text content. Profiles included in the study were gathered by querying using the website's profile search functionality with different combinations of age, gender, caste, and height.\footnote{No private data, e.g., communication, IP addresses, contact details or other user hidden data, was collected.}  Each such query returned a maximum of 600 profiles. 
While the data collected is not considered as human subjects research by our Institutional Review Board (IRB), efforts to remove personally-identifiable information were taken. All occurrences of  individuals' names in the free-form self description were removed. Unique identifiers associated with each profile such as names and usernames (after being used for de-duplication) were removed. %
We discuss in detail the ethics behind collecting this dataset and performing research on it in Section~\ref{sec:ethics}.

In total, through this method, we collected 411,292 profiles. Most profiles collected were in English (99.2\%), and the rest was filtered out using {\small \texttt{langdetect}}. 
Further, to guard against survival bias, where older profiles are potentially systematically different in attributes compared to newer profiles, we restrict our analysis to the profiles created from 2017 onward.  Also, as our topic of interest is openness to intercaste marriage in India and among the Indian diaspora in the US, we restrict the profiles to within the two countries. We filter out friend posted profiles (about 1\% of profiles) and collectively refer to profiles posted by a parent (53.5\%), sibling (43.3\%), or other relatives (3.2\%) as \textit{family-posted profiles}.  We note that although a sibling may have posted the profile, parents are typically heavily involved throughout the arrangement process on these sites \cite{seth2011online}.  In total, this filtering resulted in 313,308 profiles as shown in Table \ref{tab:data}.  

Caste is predominantly a Hindu phenomenon, with some scholars arguing that it is even intrinsic to Hinduism (for debate, see \newcite{rambachan2008caste} and \newcite{nadkarni2003caste}).  Therefore,  
we focus our research on individuals self-identifying as Hindu, which is the social context in which the effects of caste identity are strongest and most salient.  The profiles matched 352 castes, which are mapped to one of the groups shown in Table \ref{tab:castes} using the caste classification in the nationally representative Indian Human Development Survey (IHDS) \cite{desai2015india}. 
Estimating the proportion of individuals belonging to different caste categories in India is highly contentious \cite{vithayathil2018counting}, with the results on caste populations from the Socio Economic and Caste Census of 2011 (SECC) still yet to be released. 
Therefore, the population estimates for different caste categories in Table~\ref{tab:castes} are not definitive and must be treated accordingly. Broadly, relative to the available estimates of their national populations, SC and ST populations are underrepresented in our dataset, and Brahmin and OFC populations are over-represented, perhaps due to systemic economic disadvantages and differential levels of access to internet \cite{sreekumar2007cyber,Ahuja_Ostermann_2016}. The results from our analysis are also more likely to reflect urban Indians' attitudes on caste because of poor infrastructure and low Internet penetration in rural areas \cite{west2015digital}.

\section{Openness to Intercaste Marriage}
\label{sec:big-regression}

Caste remains an critical identity variable in present-day India \cite{srinivas1996caste,madheswaran2007caste,desai2012caste}. The caste of a potential spouse matters for many individuals---to the point of ruling out potential mates on the basis of this variable alone \cite{Banerjee_Duflo_Ghatak_Lafortune_2013a}.  What personal features influence whether an individual is open to intercaste marriage?  Prior studies have suggested that increased openness is associated with older age and increased education \cite{kannan1963intercaste}, being in an urban area \cite{reddy1984intercaste}, and being a member of a lower-status caste \cite{Goli_Singh_Sekher_2013}.  Yet, accurate measurements of these factors is significantly hindered by (i) cultural norms that vary substantially across states \cite{Banerjee_Duflo_Ghatak_Lafortune_2013a,Allendorf_Pandian_2016a,beteille2012peculiar} and (ii) the social desirability response bias where individuals report the more socially-acceptable option (increased openness) while privately maintaining a different attitude \cite{edwards1953relationship,paulhus1984two}. 
Here, we mitigate both issues to analyze intercaste acceptability through large-scale analysis of the features of matrimonial profiles to ask two questions: (i) to what degree do caste attitudes differ between generations and (ii) what demographic factors are associated with increased openness to intercaste marriage.

\begin{table}[t]
\centering
  \label{female}
   \resizebox{0.48\textwidth}{!}{
  \begin{tabular}{ll}
    %\toprule
    \textbf{Variable} & \textbf{Description} \\
    %\midrule
    \hline
    Gender & Male, \textbf{Female} \\
    Age & z-score of years-old\\
    Account age & log(number of days on site) \\
    First Marriage &  Yes, \textbf{No} \\
    In a Large City & Yes, \textbf{No} \\
    Education & \textbf{High school}, College, Postgraduate \\
    Income & Above Median, Below Median, \textbf{Not Answered} \\
    \makecell[l]{Parental\\~~Employment} & \makecell[l]{Both, Father, Mother, \\~~\textbf{Neither}, Not Answered} \\
    Affluence & \makecell[l]{Upper, Upper Middle Class, Lower-middle \\~~or Middle Class, \textbf{Not Answered }} \\
    Caste & Brahmin, OFC, OBC, SC, ST, Intercaste, \textbf{Other} \\
  %\bottomrule
\end{tabular}
}
\caption{Variables used in predicting openness, based on responses in online matrimonial profiles.  The reference category for all categorical variables is bolded; ages are numeric. }
\label{tab:demographic-attributes}
\end{table}

\begin{figure}[t]
    \centering
    \includegraphics[width=0.43\textwidth]{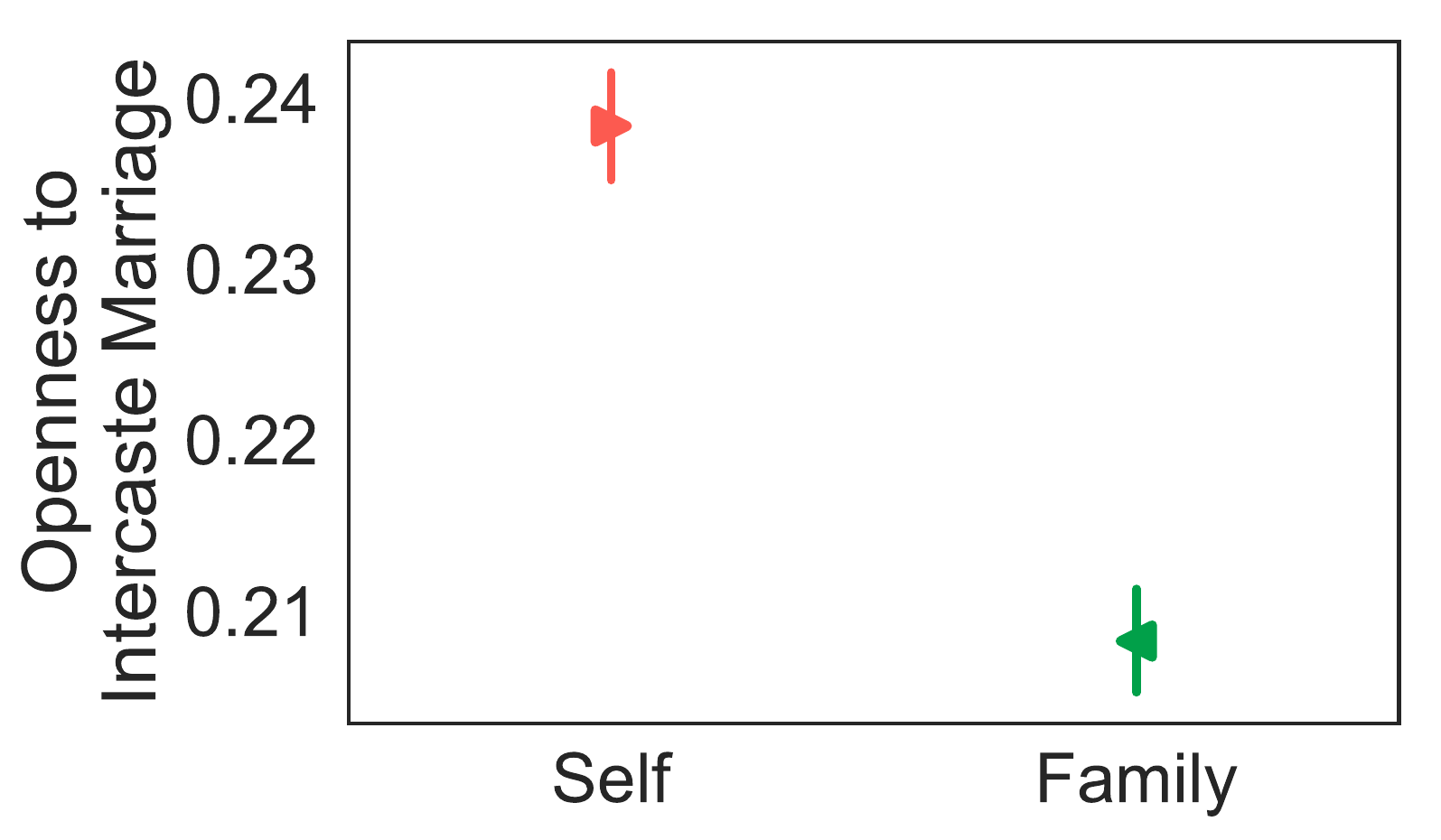}
    \caption{Under one-to-one almost exact matching conditions, profiles in which individuals posting for themselves are more open to intercaste marriages than those which were posted by family for individuals with identical demographics.  Here and throughout the paper, bars show 95\% confidence intervals.}
    \label{fig:self-vs-family-overall}
\end{figure}

\begin{figure*}[t]
\centering
  \includegraphics[width=1.0\textwidth]{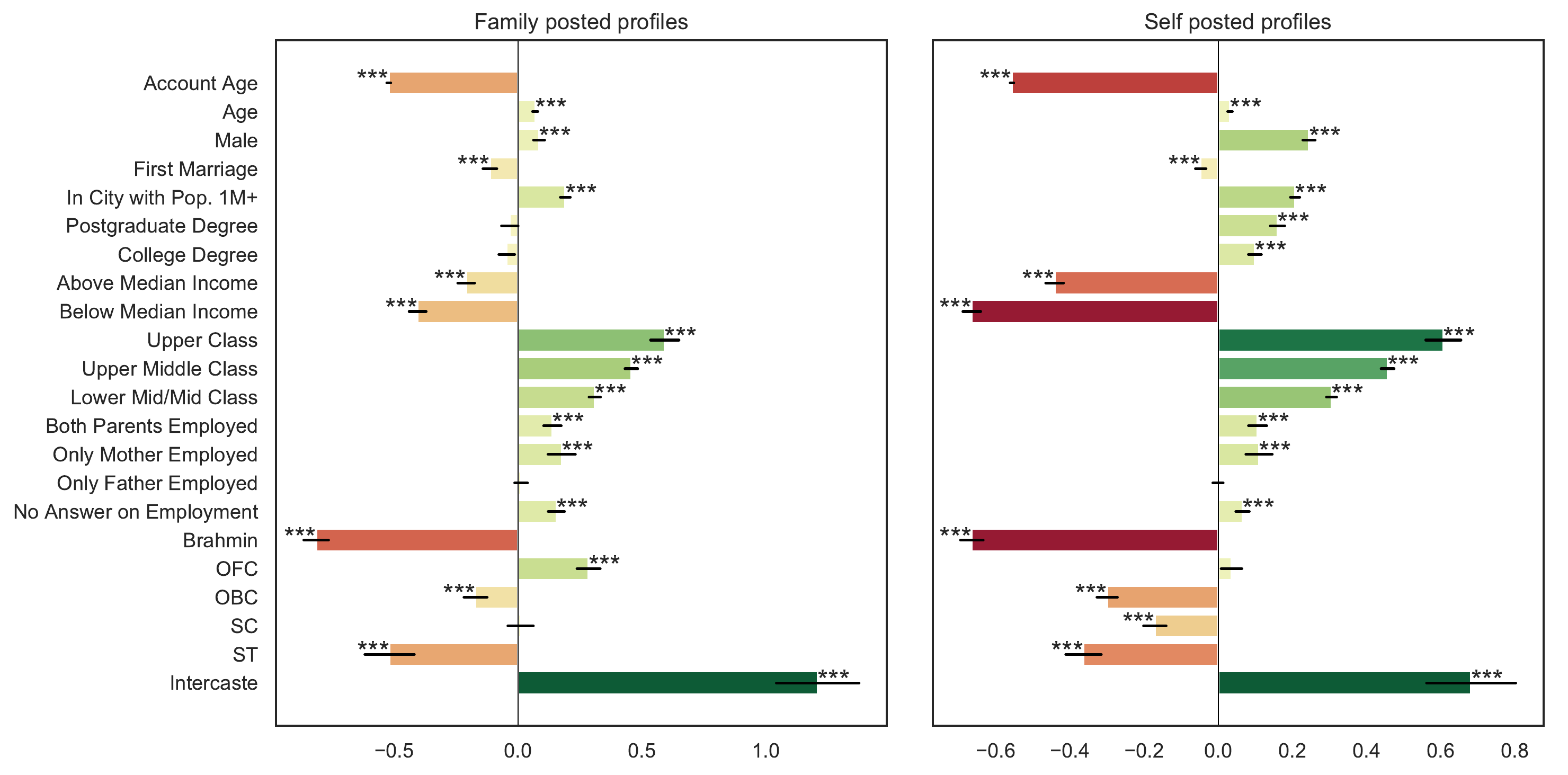}
  \caption{A comparison of demographic factors affecting openness to intercaste marriage in family-posted versus self-posted profiles shows that social status as a function of education, income, affluence, and to some degree caste, drive attitudes, where lower social status individuals are less open.
  Here and throughout the paper, significance levels for model coefficients are reported as  `***' for p$<$0.001 , `**' p$<$0.01 , and `*' p$<$0.05, and bars show standard errors. }
 \label{fig:openness}
\end{figure*}

\subsection{Data}

The website allows direct measurement of openness to intercaste marriage through its registration process where new users are instructed to indicate whether caste is an important factor for their spousal choice: ``Not particular about my partner's Caste/Sect (Caste No Bar).''  Notably, this option defaults to being \textit{not open} to intercaste marriage, underscoring the societal bias against intercaste marriage. 
While survey responses can suffer from the social desirability response bias, the direct impact of this choice on the person (i.e., marrying that potential spouse) suggests that public and private sentiments are fully aligned.  Indeed, an individual's choice directly affects how they are categorized and to whom they are visible in the websites.  Further, this choice is a blanket choice towards all members of other castes and not a response to one particular person, making it correspond more closely to a general attitude.

%\paragraph{Data construction}
These matrimonial profiles include multiple demographic attributes that are selected from a fixed set of options, shown in Table \ref{tab:demographic-attributes}.  
Affluence is a self-reported measure of socioeconomic status; income is reported in ranges or as ``decline-to-report,'' which we separate into three categories (i-ii) lower or higher than median and (iii) unknown. Although similar to affluence, income also serves as a measure of financial independence which is closely linked to intercaste marriage. 
To control for the possible effects of how long an account has been shown on the site, we include the log of the number of days, with accounts older than two years being excluded from the study altogether.
To focus solely on the attitudes on Indian residents, in this analysis, we only consider  the 298,400 India-based profiles of which 68.74\% are self-posted.

\subsection{Methods}
\label{sec:openness-model}
%\paragraph{Modeling}

Inter-generational differences in acceptance of intercaste marriage are tested through two methods. 
First, we construct control and treatment cohorts of individuals to test for differences in openness using one-to-one almost exact matching \cite{rosenbaum2010design}, which is analogous to techniques like propensity score matching except that it uses direct category alignment.  Individuals who self post are randomly paired with an individual with identical demographics (based on Table \ref{tab:demographic-attributes}) whose family had posted their profile.  To increase the percentage of individuals with counterparts, we match individuals against others with a maximum age difference of 5 years and account age difference of 90 days. Owing to our large dataset size, 76.9\% of family posted profiles are matched with a corresponding self-posted profile (versus 46.2\% with same-age, though we note that results described later are nearly identical).  Due to almost exact matching, the difference in openness is expected to be due to the identity of profile's author, i.e., the person or their family.

Second, we analyze the demographic factors contributing to increased openness by constructing identical logistic regression models for both self- and family-posted profiles using the variables in Table \ref{tab:demographic-attributes}.
Past work on caste has shown large regional differences \cite{Allendorf_Pandian_2016a}, and therefore, we include a random effect for the state that the user resides in.
To identify changes in the influence of the interest variables in self-arranged and family-arranged model of marriages, we compare log odds coefficients of the two models using the following Z test \cite{clogg1995statistical}:
%\begin{align*}
  $Z = \frac{\beta_{S} - \beta_{F}}{\sqrt{SE_{\beta_{S}}^2 + SE_{\beta_{F}}^2}} $
%\end{align*}
where $\beta_{S}$ and $\beta_{F}$ are coefficients of interest variables from both models.

\begin{figure*}[t]
    \centering
    \includegraphics[width=0.90\textwidth]{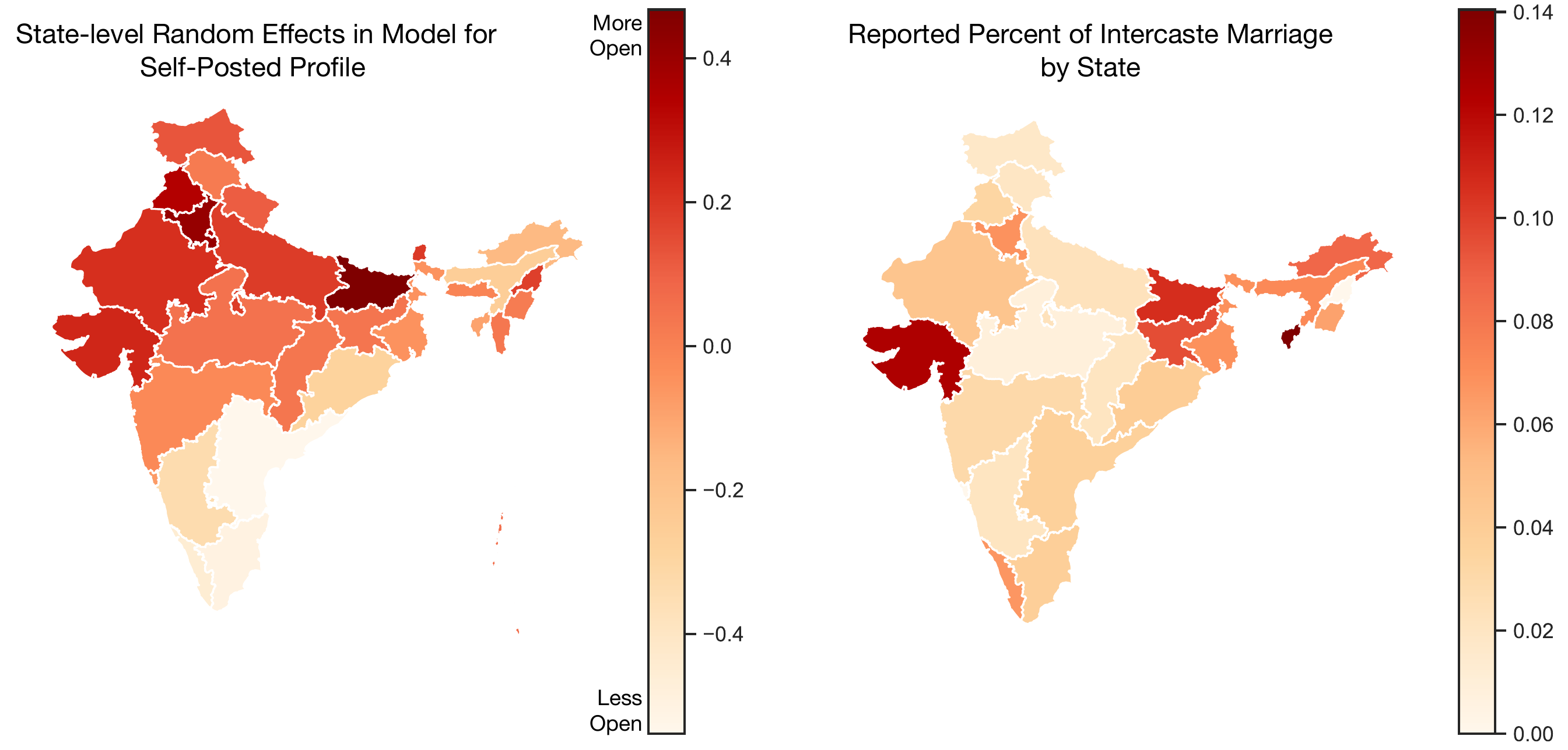} 
    \caption{A comparison of the relative openness to intercaste marriage per state, controlling for demographics, (left) with the percentage of intercaste marriages (right) reveals a stark difference in attitudes versus lived reality.  Here, openness is shown as states' random effect coefficients in the self-posted model, which controls for variation in demographic factors. We exclude two relatively less populous states, Sikkim and Meghalaya, from the right figure as their high rates of intermarriages skews visual interpretability.}
    \label{fig:geographic-trends}
\end{figure*}

\subsection{Results and Discussion}
\label{model_comparison}

Individuals are far more open to intercaste marriage than family members who write profiles for their counterparts with near identical demographics (23.94\% vs 20.82\%), as shown in Figure \ref{fig:self-vs-family-overall}.  Given that parents are heavily involved when accounts are posted by a family member, this result strongly suggests a substantial inter-generational difference in caste attitudes.  However, the current data cannot fully rule out selection effects to establish age as the only cause of this difference; individuals who opt to write their own profiles could also be more open than their counterparts having a profile written for them, thus introducing the potential for selection bias to affect the magnitude of the difference.  
Nevertheless, our result points to a substantial attitude difference in the two populations.

The demographic factors associated with openness to intercaste marriage, shown in Figure \ref{fig:openness}, suggest that a \textit{lower} social status, as a function of income, education, and affluence, is associated with less openness to intercaste marriage. 
Also, the coefficients for affluence and parents' employment are statistically indistinguishable across the two models, suggesting caste attitudes associated with these demographics are robust across generations.    
However, in the case of caste, the trend of higher status being more open deviates. Notably, the Brahmin caste is least open  to intercaste marriage dwarfing other demographic attributes.  
Together, this result identifies how various intersections of identity influence attitudes toward caste, showing that demographic variables in addition to the caste hierarchy play an important role, and pointing to the need for holistic strategies for changing attitudes, not just those based on caste.

\begin{table*}[t]
    \centering
    \begin{tabular}{ll}
        \textbf{Categories} & \textbf{Examples} \\
        \hline
        family, profession, location & Looking for someone from a good family with a settled job in Delhi\\
        caste, education & Seeking a suitable match of the same caste, ideally with a graduate degree\\
        personality & Hoping for a like-minded partner who is warm, funny and supportive\\
        education, physical attributes & Looking for a well-educated bride who is tall, fair and good-looking\\
        caste, location & Looking for a match within the same community in the US or is willing to relocate\\
    \end{tabular}
    \caption{Examples of preference statements found in profiles, lightly paraphrased to preserve privacy. }
    \label{tab:examples}
\end{table*}

Comparing both models, we identify one important difference between self- and family-posted profiles related to education and openness.   Higher education levels of the individual are positively correlated to openness in self-created profiles, whereas  a person's education level does not have a significant effect when the family creates their profiles. % 
This result suggests that family members of more-educated individuals are less supportive of their relative being in an intercaste marriage.
This infantilization of adults results in moral policing and consequently, less openness to intercaste marriage \cite{ganguly2016privacy}. However, as marriages move closer to individual choice, higher levels of education may help break caste barriers. 

Our result showing that the education of the self-poster plays a significant role in their openness directly contrasts with the result of \newcite{Ray_Chaudhuri_Sahai_2017}, who found that only the education of the husband's mother mattered but not the education of either spouse.  We attribute this difference to the fact that our analysis is able to separate out \textit{who} is initiating the search, though the family is likely to be heavily involved in the process for both \cite[p. 352]{seth2011online}.  However, our result on the impact of the mother's employment closely mirrors their result on the mother's education (both associated with increases in openness), given that employment and education are associated \cite{thomas1990intra,doss2013intrahousehold}.  Though our analyses are on slightly different data (preferences on matrimonial profiles vs. married couples), our finding presents an important insight for its implications on policy, suggesting that initiatives increasing education can potentially reduce the importance of caste in spouse selection.

Examining trends across caste identities, the relative differences in openness between castes was consistent across the models, with individuals identifying as intercaste being most open to intercaste marriage---in spite of likely being witness to and victim of their parents' social ostracism by society \cite{parish1996hierarchy}. Further, we find that in both models, openness to intercaste marriage follows the same order: Intercaste $>$ OFC $>$ SC $>$ OBC $>$ ST $>$ Brahmin. Brahmins, who occupy the highest rung in the social hierarchy, are least open to intercaste marriage.  Scheduled tribes are also resistant to intercaste marriage, which runs contrary to previous anthropological evidence \cite{sinha1962state}. Considering the small sample size, the recent Hinduization of tribal identity \cite{gautam2016hinduization}, this phenomenon should be explored further with more qualitative research.% 

\subsection{Regional variations}

Given known cultural differences across states \cite{dyson1983kinship}, when controlling for all other demographic factors, to what degree do states vary in their openness?  To answer this question, the states' random effect coefficients from the self-posted model are mapped and shown in Figure \ref{fig:geographic-trends} (left),\footnote{The map derived from family-posted accounts is highly similar and not shown here.} which reveals strong regional trends: Northern states are substantially more open to intercaste marriage than Southern states. Further, considering that the Southern states consistently rank higher in gender equity \cite{rahman2004determinants} and socioeconomic levels \cite{ravallion1999have}---variables that are associated with higher openness in our model---we would expect the states to rank higher in openness to intercaste marriage as well. This trend is likely due to strong cultural norms around cross-cousin marriages \cite{trautmann1981dravidian} and has also been observed in previous  nationally representative studies \cite{Goli_Singh_Sekher_2013}. 
Thus, our results suggest that while the \textit{prima facie} expectations of openness in Southern states might be higher, when controlling for their demographic and social differences, our results show the opposite cultural norms.
Compared to nationally representative  percentages of actual intercaste marriages shown in Figure \ref{fig:geographic-trends} (right), we also find substantial differences  between the estimated national trends of openness to actual rates of intercaste marriage.
Our result provides critical information on intercaste attitudes as relatively little information on openness to intercaste marriage is available; instead, most research relies on actual incidences of intercaste marriages, which we show differs substantially and could skew conclusions related to social inclusion. % 

\section{Attitude Shifts in Desirability}
\label{section:text}

Questions about modernization and its relation to cultural norms have long been  contentiously debated among scholars \cite{inglehart2000modernization}: Do traditional norms persist or evolve in the face of modernization? Increased economic independence and changing family dynamics in a collectivist society have been argued to lead to a more individualistic society \cite{hamamura2012cultures}. In India, the effect of globalization and urbanization has been associated with an evolution of marriage and family roles \cite{kashyap2004impact}. Given evidence of inter-generational differences in caste attitudes, does increased openness to intercaste marriage among self-posting individuals point to a shift in the attributes that are desired in a spouse, reflecting a broader shift away from caste as an important identity variable?  Here, we examine profiles' text content to test for systematic shifts in  spouse-seeking behavior.

\subsection{Data}

Many profiles include at least one statement about their desired qualities in a spouse due to a prompt from the website to include a description of who they  are looking for.  We use a regular expression to extract statements about the desired attributes of the spouse as follows.  Profiles are parsed into sentences and then filtered such that a sentence must contain a \textit{partner} word and a \textit{search} verb; here,  search-related verbs are seek, need, search, look, looking, want, prefer, desire, expect and wish; and partner-related words include person, hubby, spouse, individual, soul mate, soulmate, partner, someone, some one, alliance, match, prospective and prospect.
To streamline profile creating, the website provides auto-generated descriptions that users can copy and use, e.g., ``I am seeking someone who will be a great partner in my journey of life.''  To remove  possible effects from auto-generated text, we filter out any sentence repeated verbatim over 10 times from the analysis.
Through this approach we identified sentences containing partner preferences from 55,520 profiles with a median length of 73 characters. Table \ref{tab:examples} lists  paraphrased examples of extracted sentences.

\subsection{Methods}

To test for differences in spousal preference, we mirror prior work in testing for mentions of specific categories such as appearance \cite{perilloux2011meet}. Past work in economics on dowry determinants suggested that education, height, and older men were preferred \cite{Dalmia_2004a}. Other evolutionary psychology work suggests that parents look for good character, family background, similar social status, wealth, health and chastity \cite{Apostolou_2010}. Work on partner preferences in India highlighted the importance of kindness and understanding, health, mutual attraction, education and intelligence for males and females \cite{Kamble_Shackelford_Pham_Buss_2014}. Men preferred young, physically attractive and house work oriented partners while women preferred attributes such as high social status, education, income and ambition \cite{Kamble_Shackelford_Pham_Buss_2014,Prakash_Singh_2014}. 
Using these theory-based themes as reference, two annotators independently classified profile words into theory driven categories. % 
Manual coding was done to account for emerging themes such as ``location'' which was not predicted by past research.  After the first round of coding, discussions on word classifications and new themes were conducted before recoding, reaching high agreement at 0.91 Krippendorff's $\alpha$.  Categories and example words are shown in Table \ref{tab:preference-categories}. These words are matched to the extracted preference statements in self-posted profiles.

\begin{table}[t]
    \centering
    \begin{tabular}{ll}
        \textbf{Category} & \textbf{Example words} \\
        \hline
        Age & years, young, younger\\
        Attractiveness & beautiful, handsome, pretty\\
        Caste & community, brahmin, iyer, \\
        Education & well-educated, degree, IIT \\
        Family & parents, mom, families\\
        Finances & earning, financially, career\\
        Health & smoking, non-smoker, teetotaler\\
        Location & USA, mumbai, relocate\\
        Personality & loyalty, honesty, funny\\
        Physical attributes & slim, tall, fair\\
        Profession & profession,  work, job
        
    \end{tabular}
    \caption{Categories of words for aspects of a desired spouse. }
    \label{tab:preference-categories}
\end{table}

\subsection{Results and Discussion}

\begin{figure}
    \centering
    \includegraphics[width=0.47\textwidth]{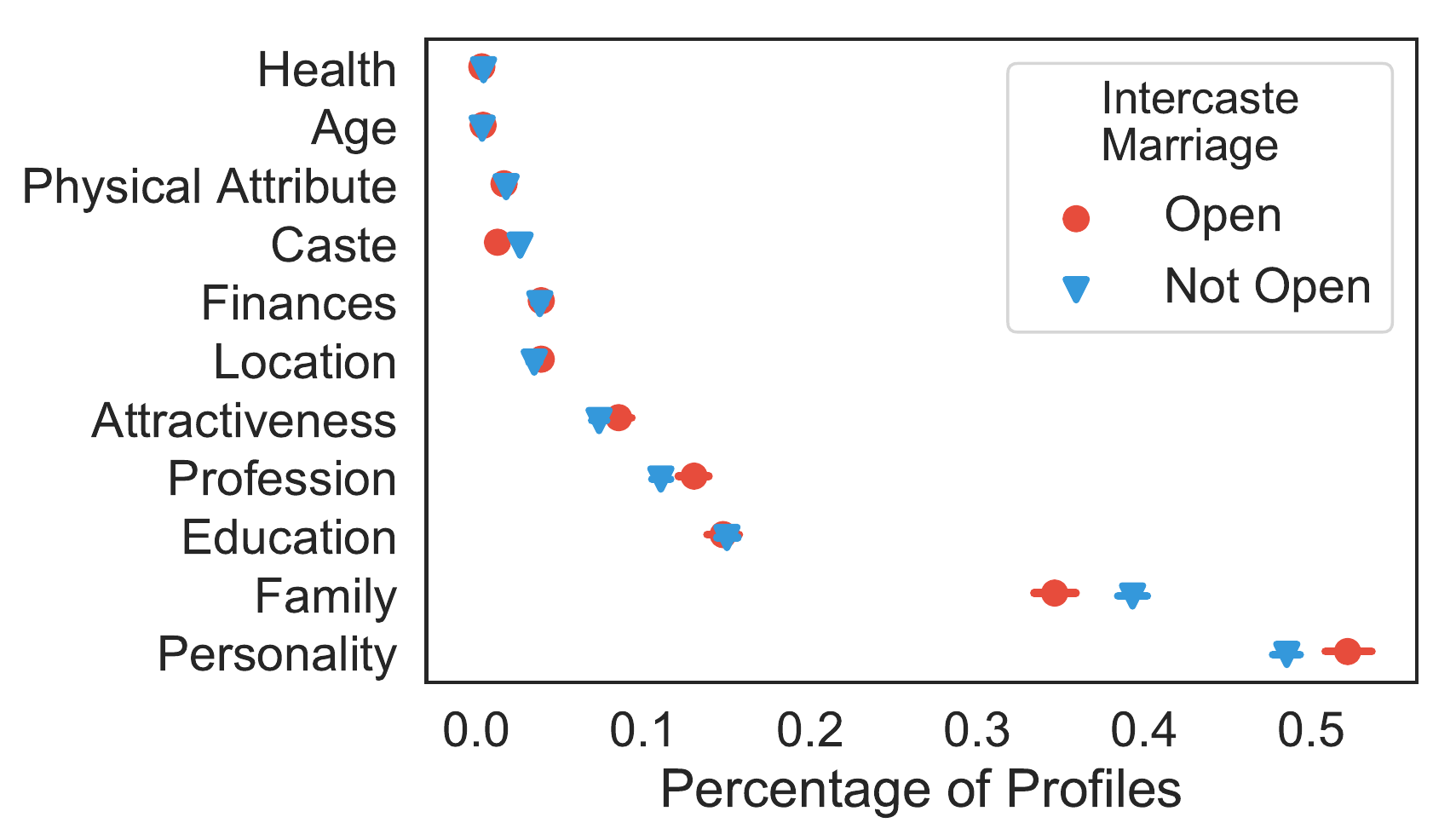}
    \caption{The percent of self-posted profiles mentioning categories of desired qualities shows individuals who are open to intercaste marriage are more likely to seek spouses on the basis their personality, rather than family background. }
    \label{fig:profile-usages}
\end{figure}

Both individuals who are open to intercaste marriage and those who are not have highly similar preferences in the attributes they look for in a spouse, as shown in Figure \ref{fig:profile-usages}.  This similarity implies that individuals open to intercaste marriage are no less choosy in the spousal choice.  
However, the relative mentions of Profession, Personality, Caste and Family significantly differed between the groups (p$<$0.01), with the largest differences in the two most-frequently mentioned categories, Family and Personality. Specifically, we find that individuals open to intercaste marriage use a higher frequency of words related to the potential spouse's personal qualities, whereas individuals who are not open use words more related to the social context of the spouse (caste and family).  Individuals preferring to specify specific personal qualities as opposed to caste or family background in stating their preferences is a step forward in caste attitudes. A lower emphasis on family background of the prospective spouse suggests a turn towards a more individualistic concept of marriage, moving away from the traditional notion of marriage being a union of two families.
Our large-scale data-driven results indicating a shift from family to personality provide further grounding to the qualitative findings from in-person interviews of individual's preferences when seeking spouses on these sites \cite{titzmann2013changing}.

\section{Openness to Caste in Diaspora}
\label{sec:diaspora}

Economic and social mobility have led to substantial populations of Indians living abroad, both as immigrants and as citizens.  Assimilation and modernization theories suggest that when immigrants move to more individualistic societies, they adopt individualism and the family plays a smaller role in individuals' lives \cite{nee2012rethinking,tipps1973modernization}.
However, the bi-cultural integration model of acculturation \cite{maisuria2004growing} contends that migrants retain cultural values in private settings while inculcating the host culture in professional settings. 
Using over 14,000 matrimonial profiles of the Indian diaspora living in the US, we test to assess whether attitudes and demographic factors associated with openness persist across cultural settings.  

This study is rooted in a series of anthropological investigations of whether immigrants carry caste to their adopted lands.  Studies on the Indian diaspora in the Caribbean islands report a decline and elimination of caste consciousness \cite{roopnarine2006indo}.  However, similar research in the US demonstrates that caste and casteism continues to be prevalent among the Indian diaspora \cite{adur2017stories,maira2012desis}.  A crucial difference between the groups in the Caribbean and US is that the former were low-caste indentured workers with little cultural capital \cite{roopnarine2003east}, while immigrants in the US are mostly high skilled professionals such as engineers and doctors.  Thus, understanding factors that influence intercaste marriage in the US provides insight into how traditional caste boundaries are erased or reinforced in a relatively-privileged minority community within a larger western context.

\subsection{Data and Methods}
We analyze data from 14,908 matrimonial profiles created by US-based users. First, we compare US based individuals to their Indian counterparts on openness to intercaste marriage. Then, we examine how demographic factors influence their openness to intercaste marriage. 
US and Indian profiles are compared using the one-to-one almost exact matching technique outlined in Section \ref{sec:openness-model}, matching users based on all demographic attributes in Table \ref{tab:demographic-attributes}. 89.6\% of US based individuals are matched with a random counterpart in India with identical demographics.

We repeat a similar setup as in Section \ref{sec:openness-model}, fitting a random effects logistic regression model on openness as the dependent variable.  Here, we include the US state as a random effect and include an additional fixed effect for whether the person was raised in the US. Because of fewer profiles available for this analysis, we train a single model using both self- and family-posted profiles and include a fixed effect for whether the profile is self-posted. Based on results from Section \ref{model_comparison} which showed factors affecting openness in self- and family-posted profiles are similar, we do not expect that using a single model will affect validity.

\begin{figure}[t]
    \centering
    \includegraphics[width=0.43\textwidth]{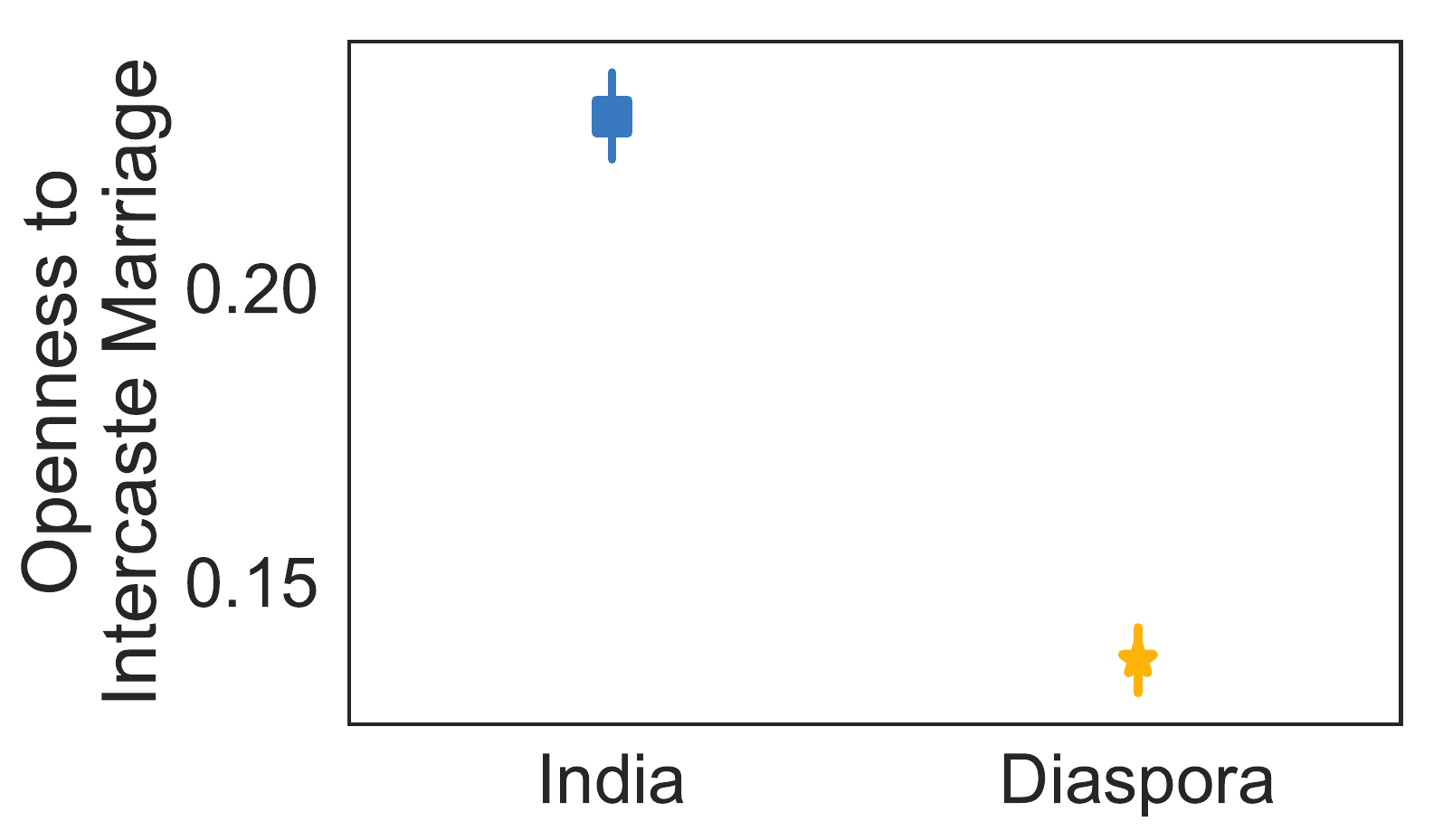}
    \caption{Under one-to-one almost exact matching conditions, individuals in the USA posting for themselves are less open to intercaste marriages than their counterparts with identical demographics in India.}
    \label{fig:usa-vs-india-overall}
\end{figure}

\subsection{Results and Discussion}

\begin{figure}[t]
  % Note: text width is the width of both columns
  \includegraphics[width=0.48\textwidth]{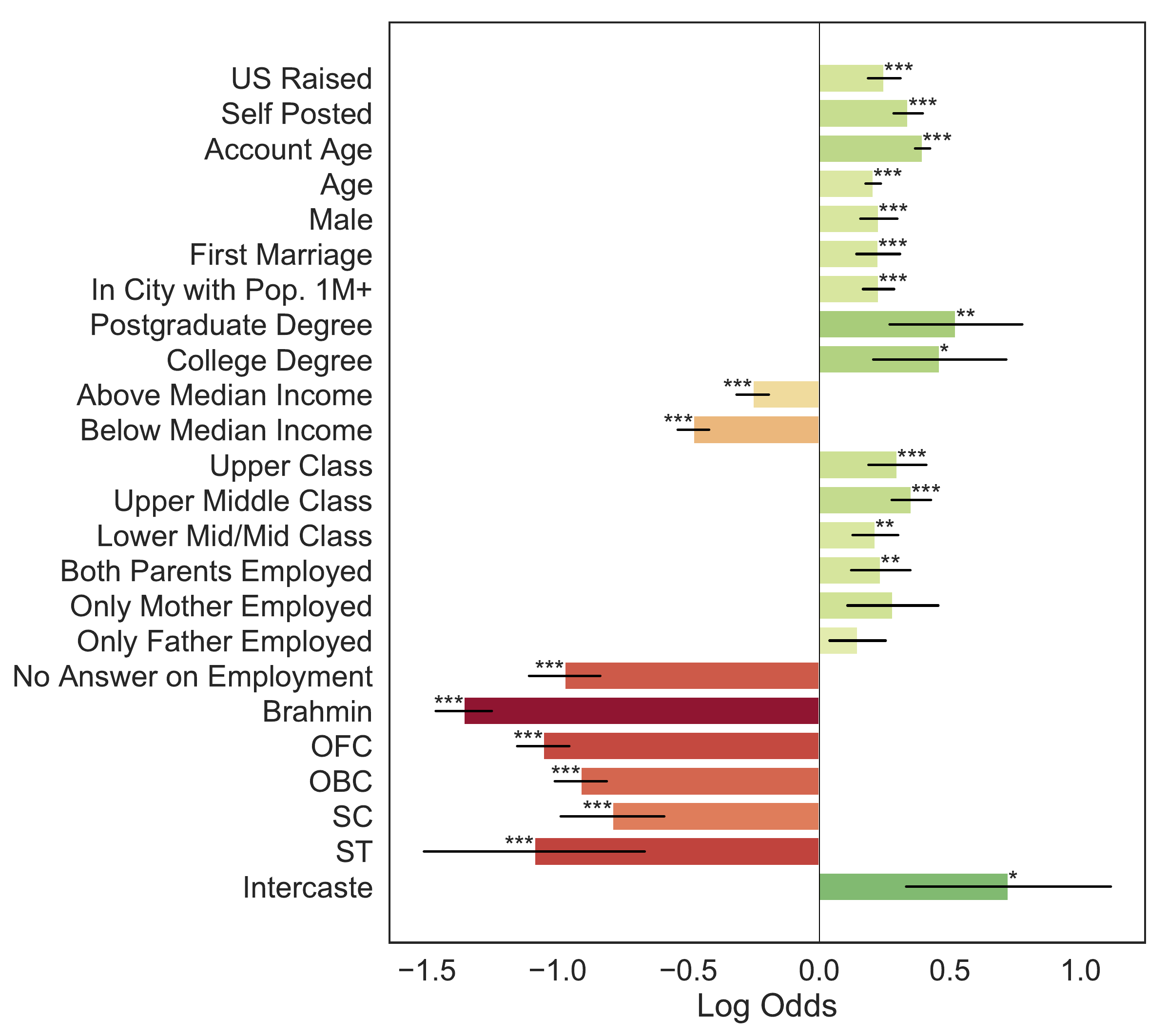}
  \caption{Demographic factors that influence openness to intercaste marriage in the diaspora, shown as their regression coefficients. These effects are largely similar to Indian residents (cf. Figure \ref{fig:openness}).
  }
  \label{fig:diaspora}  
\end{figure}

Comparing the matched profiles in Figure \ref{fig:usa-vs-india-overall}, we find that individuals in the US are much less open to intercaste marriage than individuals with identical demographics in India (13.69\% vs 22.91\%).  This result supports  theory from social psychology regarding South Asian immigrants that suggests immigrants construct an essentialist notion of ethnic identity for self protection, which valorizes their culture and, in this circumstance, preserves caste salience \cite{mahalingam2013cultural}. 

The demographic aspects associated with increased openness to intercaste marriage in US diaspora, shown in Figure \ref{fig:diaspora}, largely mirror those of India-based profiles (cf. Figure \ref{fig:openness}).  This result matches the expectations of the bi-cultural integration model of acculturation \cite{maisuria2004growing} in which caste preferences are maintained in the more private setting of Indian-specific matrimonial sites, relative to the larger cultural (American) attitudes about caste.  However, compared with Indian immigrants, US-raised Indians are \textit{more} open to intercaste marriage, which supports the modernization theory that individuals will adopt aspects of the surrounding environment. Modernization theory notwithstanding, we find that the positive openness effect from being US raised is dwarfed by the resistance based on status as a function of income and the caste hierarchy---underscoring the importance of recognizing intersectional identity in understanding caste attitudes. 

Similar to trends in India, the demographics of self-posting, age, residing in a large city, more education, and higher income---all signals of higher social status---are positively linked to openness. 
Contrary to trends in Section \ref{model_comparison}, individuals are more open to intercaste marriage during their first marriage, which should be more deeply explored. We speculate that individuals looking to get remarried may have married outside caste or ethnicity in the previous marriage, and are looking to seek in-group affinity within their own community. However, more data on their previous marriage is required to make definitive conclusions.

\subsection{Openness and Opportunity}

Indian diaspora have settled throughout the US within varying sizes of communities.  Prior studies on mate selection has found that with increased access to potential spouses, individuals become more selective \cite{south1991sociodemographic}.  However, prior sociological studies have also shown that urbanness is associated with increased tolerance for others of different races and background \cite{fischer1976urban}.  Given increased access to a larger pool of potential spouses who may also belong to the same caste, are individuals less open to intercaste marriage (i.e., more selective)?

\begin{figure}
    \centering
    \includegraphics[width=0.40\textwidth]{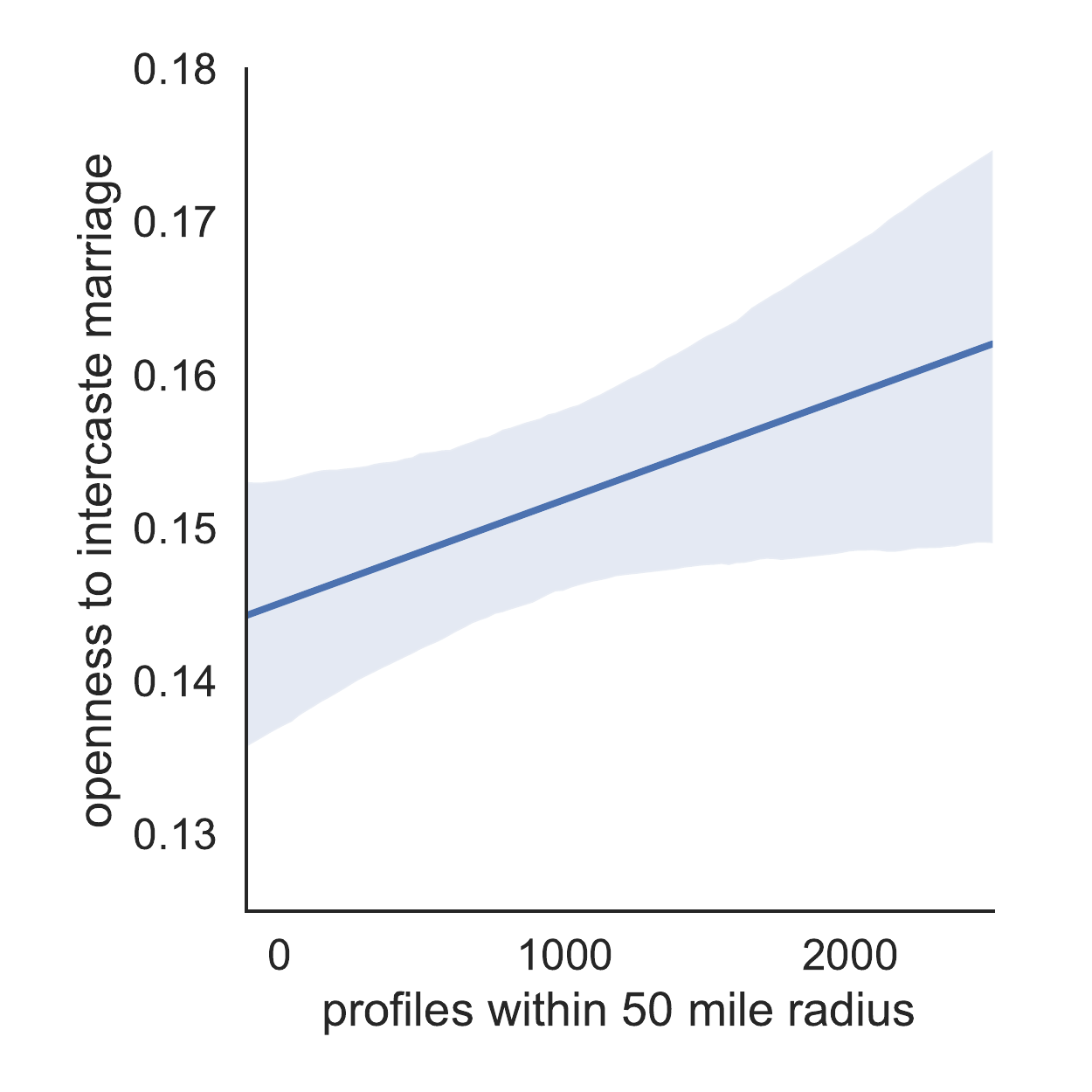}
    \vspace{-5mm}
    \caption{Openness to intercaste marriages is positively associated with the number of potential spouses nearby, shown here as the number of profiles for people within 50 miles. Shaded regions denote the 95\% confidence interval for a logistic regression fit using the number of nearby profiles.  }
    \label{fig:openness-vs-density}
\end{figure}

To test the hypothesized relationship between openness and opportunity for marriage, we map the locations for all profiles and compute the number of profiles within a 50 mile radius as a proxy for potential matrimonial opportunity.  The number of profiles correlates highly with the nearby urban population ($r$=0.90) and therefore also serves as a proxy for urbanness. 
A logistic regression is then fit for predicting a profile's openness from the number of profiles.  The results, shown in Figure \ref{fig:openness-vs-density}, reveal an \textit{increase} in openness as the number of potential spouses increase.  This result suggests that increasingly urban settings are associated with less discriminatory attitudes towards caste.  However, we note that this current evaluation in insufficient to establish causality; indeed, the effect could be due to a selection bias where more open individuals move towards urban areas.  Nevertheless, the results do indicate that despite increased opportunities to marry within their own caste, individuals in larger cities increasingly choose to be open to intercaste marriages.

\section{Ethical Considerations}
\label{sec:ethics}

The use of matrimonial profile data in research warrants a discussion of the ethical implications and decisions made.  The approach taken in this study is especially influenced by recent work around ethical research using social media data \cite{zimmer2017internet,vitak2016beyond,fiesler2018participant}.  In particular, precautions were taken to ensure the design, decision-making, data collection, and handling was done in an ethical manner according to recommended best practices and guidelines \cite{townsend2016social,Mislove_Wilson_2018,markham2012ethical}.
Here, we discuss three key aspects: (1) whether this data is public, (2) data and privacy protection practices, and (3) the overall risk-benefit trade-off in analyzing the data.

\myparagraph{Public or Private}

Research using public information on the internet is typically not considered human subjects research  \cite{vitak2016beyond}.  The IRB at the authors' home institution of the University of Michigan also verbally confirmed that they do not consider work on this data to be human subjects research because all the information analyzed is public.  However, the IRB is primarily concerned with legal and institutional regulations, and therefore, we consider the ethical implications of considering this data to be public.  In their discussion of the  process for deciding whether using particular data is ethical, \newcite{townsend2016social} highlight a case study on data from a dating app that is a close analog to our study.  Here and in the case study, individuals posting the data have the expectation that strangers will be viewing their profiles, and therefore we consider the data to be public in nature.  However, as users do not expect their profiles to be publicly indexed and archived outside the site, releasing the data may violate their contextual expectations of privacy \cite{nissenbaum2011contextual}. Also, previous releases of anonymized data have been de-anonymized, putting users at possible risk \cite{zimmer2010but}. Therefore, we report only paraphrased examples and, though it limits reproducibility, opt not to release the data.

\myparagraph{Data Privacy and Protection}
Although we consider the data to be public, substantial precautions were taken during the data collection process to ensure that no personally identifiable information was retained.  We collect only the minimal data required to study the relationship of personal attributes associated with caste discrimination (e.g., education, caste identity).  This process intentionally excludes data such as  photos, contact details, their interactions with other users, or users' listed preference in other partner attributes, apart from openness to intercaste marriage.  Once the data collection process was finished, all unique identifiers associated with each profile such as names and usernames (after being used for de-duplication) were removed.  The fully-anonymized profiles are then stored on an encrypted hard disk accessible only to the researchers involved in the study.  %
Although our data collection process required the creation of a new account, no communication or contact was made with any users on the site before, during, or after collection.

\myparagraph{Risk-Benefit Analysis}
In critically examining this project, we aim to minimize potential harm while maximizing the societal benefits of our research towards understanding casteism.  
Casteism is a serious and widespread form of social discrimination that results in major emotional, financial, and, at times, physical harm to those experiencing it, as detailed in Section~\ref{sec:caste}. Our research provides valuable insights into where efforts to mitigate these discriminatory attitudes might be best placed; for example, geographic findings (Figure \ref{fig:geographic-trends}) show a stark difference in geographic trends in our data when compared to the limited %but best-available 
data on rates of intercaste marriage, which point to where governmental and NGO efforts might be better addressed.  As another example, while  \newcite{Ray_Chaudhuri_Sahai_2017} found no association between discriminatory attitudes and education levels of the married couple, our regression analysis (Figure \ref{fig:openness}) shows that increased education is in fact associated with decreased discriminatory attitudes in self-posted profiles. Our research now identifies education as a potential long-term intervention strategy to change cultural attitudes  in certain situations.  

The largest risk from this research is loss of individuals' privacy.  As a result, we have attempted to mitigate privacy concerns by anonymizing the collected data and by our decision to not share the data.  We report only general trends---indeed, these general trends are what are most critical for directing efforts towards removing the scourge of casteism.  

Finally, while the website's Terms of Service (TOS) prohibit crawling, our choice to disregard these terms was again motivated by a harm-benefit analysis.  Our crawler issued limited queries that were appropriately spaced to have minimal impact on the site and our data privacy practices minimize the potential privacy risks to users. Our research from the resulting data presents an unprecedented view into caste attitudes that is otherwise difficult to measure and thus provides a substantial benefit towards understanding caste attitudes and their prevalence.  This harm-benefit reasoning is analogous to those used by audit studies studying discrimination on platforms whose TOS prohibit such activities \cite{hannak2017bias,chen2018investigating}.

\section{Conclusion}

Caste based discriminatory attitudes currently stigmatize millions of individuals, denying the equal access to employment, education, and even basic human rights. Governmental incentives and social movements have attempted to counter these attitudes, yet accurate measurements of public opinions on caste are unavailable despite being essential for understanding whether progress is being made.  Here, we introduce a novel approach to measuring public attitudes of caste through an indicator variable: openness to intercaste marriage.  Using a dataset of over 313K profiles from a major Indian matrimonial site, we precisely quantify attitudes on intercaste marriage, along with differences between generations and between Indian residents and diaspora.

Our work provides the following three main contributions towards understanding attitudes towards caste.
First, we show attitudes are changing between generations, with younger individuals being more open to intercaste marriage and, in a holistic analysis of identity, show that lower social status as a function of multiple factors (e.g. income, education) is  predictive of decreased openness to intercaste marriage.
Further, we provide the first large-scale measurement of openness to intercaste marriage, showing that attitudes are not well aligned with incidences of intercaste marriage.
Second, we uncover signs of cultural shift towards individualism by examining the desired qualities in a spouse: as attitudes become more liberal towards intercaste marriage, less emphasis is on the familial aspects of a spouse and, instead, more emphasis is on their individual personality.
Finally, we find that Indian diaspora are substantially less open to intercaste marriage.  While some selection bias exists for diasporic individuals seeking a spouse on an Indian matrimonial site, even when controlling for where they grew up, our results support the bi-cultural theory of integration \cite{maisuria2004growing} where individuals adopt the norms of the host country in public settings while maintaining their own culture's norms in private settings. % 
Our research provides the first empirical evidence identifying how various intersections of identity shape attitudes toward intercaste marriage in India and among the Indian diaspora in the US, where caste hierarchy, transnational social location, social status, and other demographic variables all play an important role.

\section*{Acknowledgements}

The authors thank Ceren Budak, Paul Resnick, and the Computational Social Science seminar at UMSI for their helpful feedback on earlier drafts of this work.

\fontsize{9.8pt}{10.8pt} \selectfont

\bibliography{ms}
\bibliographystyle{aaai}

\end{document}